\documentclass[english,floatfix,twocolumn,superscriptaddress,prl,aps]{revtex4}
\usepackage[T1]{fontenc}
\usepackage[latin1]{inputenc}
\usepackage{graphicx}
\usepackage{amssymb}
\usepackage{amsmath}
\usepackage{bm}
\usepackage{babel}
\usepackage{color}
\makeatother

\begin{document}

\title{Making Graphene Luminescent}

\author{T. Gokus$^1$, R.R.Nair$^2$,A.Bonetti$^3$, M. B{\"o}hmler$^1$, A.Lombardo$^3$,\\K. S. Novoselov$^2$, A.K. Geim$^2$, A. C. Ferrari$^3$\footnote{acf26@eng.cam.ac.uk},A. Hartschuh$^1$}
\affiliation{$^1$Chemistry and Biochemistry Department and CeNS,Ludwig-Maximilians- University of Munich, Germany\\
$^2$Department of Physics and Astronomy, Manchester University, UK\\
$^3$Engineering Department, Cambridge University, Cambridge, UK}

\begin{abstract}We show that strong photoluminescence (PL) can be induced in single-layer graphene on using an oxygen plasma treatment. PL characteristics are spatially uniform across the flakes and connected to elastic scattering spectra distinctly different from those of gapless pristine graphene. Oxygen plasma can be used to selectively convert the topmost layer when multi-layer samples are treated.\end{abstract}
\maketitle

Graphene is at the center of a significant research effort\cite{Nov306(2004)}. Near-ballistic transport at room temperature and high mobility\cite{zhang,mozorov,andrei} make it a potential material for nanoelectronics\cite{Han,Chen}, especially for high frequency applications\cite{Yuming}. Furthermore, its transparency and mechanical properties are ideal for micro and nanomechanical systems, thin-film transistors, transparent and conductive composites and electrodes\cite{bunch,hernandez,eda}, and photonics\cite{zhipcondmat}. There are two main avenues to modify the electronic structure of graphene. One is by cutting it into ribbons and quantum dots\cite{Han,Chen,Dai_SCI08,stamp1,pono}, the other is by means of chemical or physical treatments with different gases, to reduce the connectivity of the $\pi$ electrons network\cite{elias,ruoff09}. One of the most popular insulating chemical derivatives is graphene oxide (GO)\cite{ruoff09}. Bulk GO solutions and solids do also show a broad luminescence background\cite{daigo,kikkawa}. Hydrogen plasma was used to controllably and reversibly modulate the electronic properties of individual graphene flakes, turning them into insulators\cite{elias}. Aggressive oxygen treatment was applied to create graphene islands\cite{flynn08}. However, thus far, no photoluminescence (PL) was seen from individual graphene layers, either cut into ribbons or dots, or chemically treated, making graphene integration into optoelectronics still elusive.

Graphene samples are produced by micro-cleavage of graphite on a silicon substrate covered with 100 nm SiO$_2$\cite{Nov306(2004)}. The number of layers is determined by a combination of optical microscopy and Raman spectroscopy\cite{ACFRaman,CasiraghiNL}. Optical imaging at 473 and 514nm is done in an inverted confocal microscope. The beam is reflected by a splitter and focused with a high numerical aperture objective (NA=0.95). Raman spectra are measured at 514nm with a Renishaw micro-Raman spectrometer. The samples are then exposed to oxygen/argon (1:2) RF plasma (0.04mbar, 10W) for increasing time (1 to 6 seconds). The structural and optical changes are monitored by Raman spectroscopy and elastic light scattering. PL decay dynamics is recorded by time-correlated single photon counting (TCSPC) upon pulsed excitation at 530 nm (2.34 eV), with a time-resolution of $\pm$3ps. The acquisition time per pixel is of the order of few tens ms. The spatial resolution is $\sim$800nm. The power on the sample is well below 1mW, to prevent photo-damage.
\begin{figure}
\centerline{\includegraphics [width=90mm]{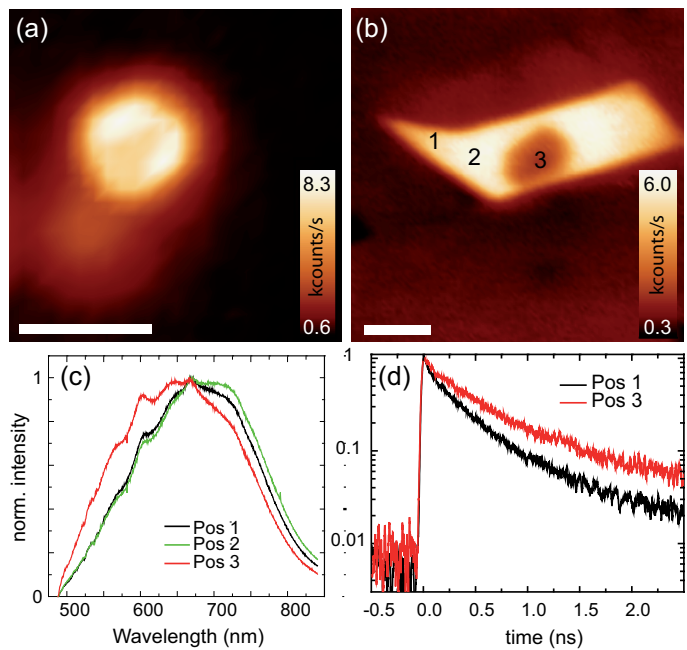}}
\caption{\label{Fig_image}(a) Confocal PL image excited at 473nm (2.62eV) for a graphene sample oxidized for 3s. Scale bar 5$\mu$m. The bright PL spots are spatially localized. (b) Uniform emission after 5s. Scale bar 10$\mu$m. For position 3 in (b), PL is bleached intentionally by intense laser irradiation. (c) Spectra detected at the positions marked in (b). These have broad PL centered $\sim$700nm (1.77eV). (d) PL transients detected at the positions indicated in (b). The dynamics can be described by a triple-exponential with decay times $\sim$40ps, 200ps and 1000ps.}
\end{figure}

Spatially resolved PL shows bright and localized emission for short treatment times(Fig.~\ref{Fig_image}(a)). For longer times, the PL is strong and spatially homogeneous (Fig.~\ref{Fig_image}(b)), with a single broad band centered at $\sim$700nm (1.77eV), Fig.~\ref{Fig_image}(c). Intense laser excitation with power exceeding 1mW leads to photo-bleaching and a PL maximum blue-shift. The excited state decay dynamics of this photoluminecent graphene (PLG) is complex. The PL transients of Fig.~\ref{Fig_image}(d) can be described by a three-exponential decay with lifetimes $\sim$40, 200,1000ps, substantially longer than those observed in semiconducting nanotubes and amorphous carbon\cite{gokus,ley}. Remarkably, the PL transients are nearly uniform across the complete spectrum. This implies that spectral diffusion due to energy migration, typical for heterogeneously broadened systems, is absent (see Figure 6 in Methods).

A Raman investigation gives further insights into the evolution from pristine graphene to PLG. Fig. 2 plots the Raman spectra and the main fitting parameters (see Methods). A broad PL background is evident in Fig 2a for treatment times above 2s. This is quite different from the case of the hydrogen plasma treated samples of Ref.\cite{elias}, where no luminescence was observed. Fig 2a also shows a significant increase of the D and D' intensities, and the D+D' combination mode $\sim$2950 cm$^{-1}$, which requires a defect for its activation. Note that in defected graphene the relaxation of the backscattering condition results in significant broadening of the second order modes. Defect scattering also broadens the first order peaks, eventually merging G and D' in a single large G band for treatment times above 1s.
\begin{figure}
\centerline{\includegraphics [width=95mm]{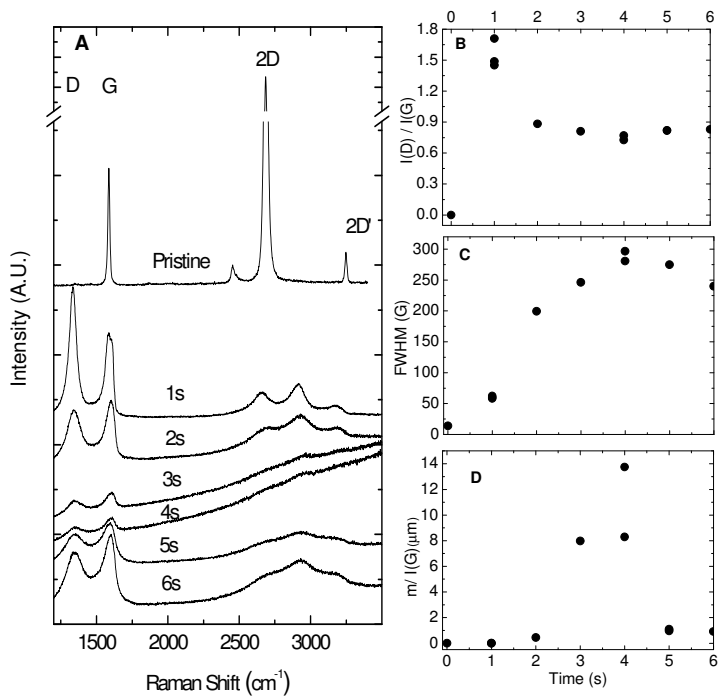}}
\caption{\label{FigRaman}Raman characterization as a function of treatment time. (a)Raman spectra;(b) D to G intensity ratio;(c) FWHM(G); (d) ratio of the slope of PL background (m) to the G peak intensity, I(G)}
\end{figure}

Fig 2b plots the evolution of the D to G peak intensity ratio, I(D)/I(G). This first raises and then decreases for increasing time. The D peak intensity is a measure of the number of defects\cite{Ferrari00,tk} (see Methods). Ref.\cite{tk}, noted that I(D)/I(G) varies inversely with the cluster size $L_a$ in poly- and nano- crystalline graphites: I(D)/I(G)=C($\lambda$)/L$_a$, where C(514.5nm)$\sim$4.4 nm from Refs.\cite{tk,knight}. This is known as the Tuinstra and Koening relation (TK). TK holds until a critical defect density. Since the D peak requires the presence of sixfold rings, when the network starts losing them, I(D) decreases with decreasing $L_a$\cite{Ferrari00}. In this case I(D)/I(G)=C'($\lambda$)L$_a$$^2$, with C'(514.5nm)$\sim$0.55nm$^{-2}$\cite{Ferrari00}. Combining the latter with TK, we deduce that Fig.2b shows a continuous L$_a$ decrease down to$\sim$1nm, and a transition to a network with fewer sixfold rings for treatment longer than 1s. This is further validated by considering the evolution of the Full Width at Half Maximum of the G peak, FWHM(G). In defect-free graphene, a variation of FWHM(G) is observed as a consequence of doping\cite{cinzapl,das,pisana}. However, in the case of defected samples, peak broadening is a result of the activation of $\textbf{q}$$\neq$$\textbf{0}$ phonons. An empirical correlation between FWHM(G) and $L_a$ was reported in Ref.\cite{prbcn} considering a variety of disordered and amorphous carbons. Comparing FWHM(G) in Fig.2c with the trend in Ref.\cite{prbcn}, again we get L$_a\sim$1nm for the longest treatment. The large FWHM(G) also implies a distribution of L$_a$ around the average value.
\begin{figure}
\centerline{\includegraphics [width=80mm]{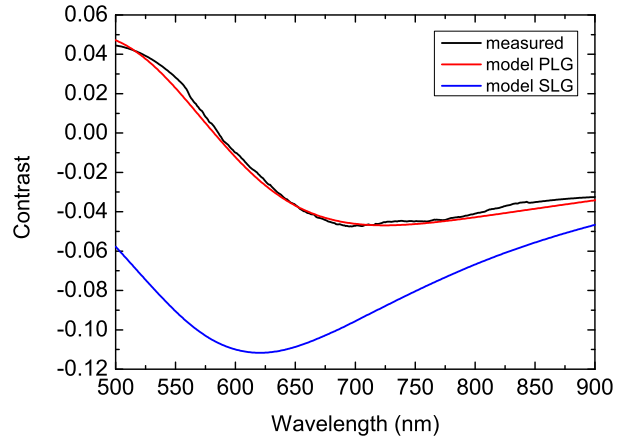}}
\caption{\label{Fig_scattering} White-light spectrum of PLG (black line) compared to pristine graphene (blue Line)\cite{CasiraghiNL}. The red line is a PLG model using a Cauchy function for the complex refractive index (see text)~\cite{ruoff08}.}\end{figure}

The ratio of the slope of the PL background (m), to I(G) is often used in disordered carbons as a measure of the PL strength, when comparing different samples\cite{cinzprb}. We thus plot m/I(G) in Fig 2(d). This reaches a maximum, then decreases for the longest treatment, consistent with the lack of PL in Ref.\cite{flynn08} after oxygen treatment targeted at layer removing.
\begin{figure}
\centerline{\includegraphics [width=80mm]{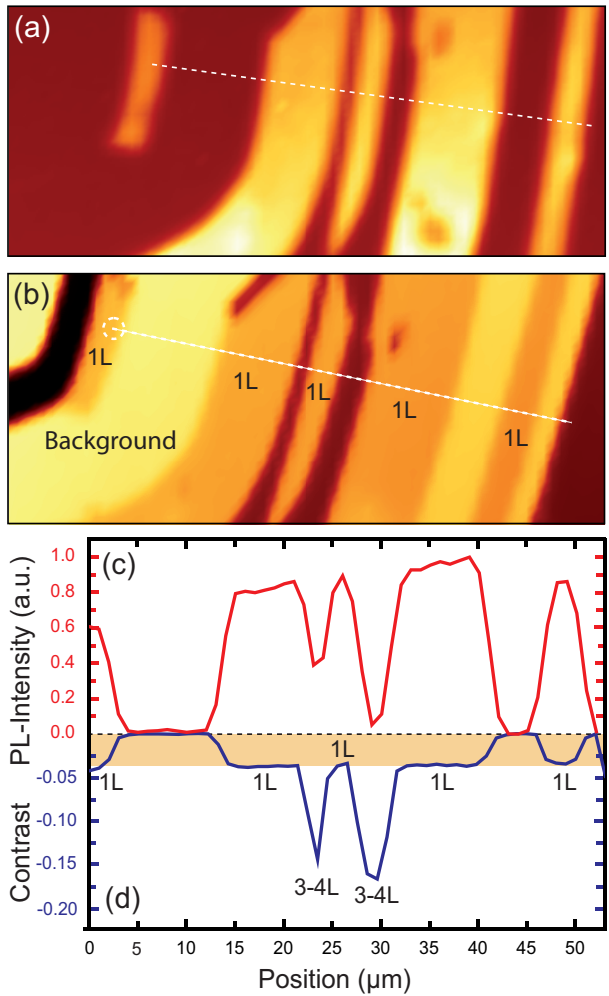}}
\caption{\label{Fig_PL-scatt}Correlation between PL and layer thickness.(a)PL image;(b)scattering image of the same sample area.(c, d)Corresponding cross sections taken along the dashed lines in (a,b). PL is only observed from treated SLG, marked 1L}
\end{figure}

White-light scattering spectroscopy of PLG reveals clear differences to pristine graphene. Fig.~\ref{Fig_scattering} illustrates the scattering
spectra of both materials on SiO$_2$/Si. While pristine graphene appears dark throughout the spectral range covered in the experiment, corresponding to
a negative interferometric contrast~\cite{CasiraghiNL}, PLG shows weaker contrast with a positive sign for wavelengths smaller than 580nm. The spectrum of PLG yields the complex refractive index $n'=A_n+B_n/\lambda^2 +i*(A_k+B_k/\lambda^2)$, with Cauchy parameters $A_n$=2.76, $A_k$=0.06, $B_n$=3000, $B_k$=1500 for the graphene thickness of 0.34nm, comparable to those obtained for GO~\cite{ruoff08}.

The data presented so far are taken for single layer graphene (SLG). A different behavior is observed for multi-layer graphene (MLG), which remains non-luminescent following treatment. Indeed, PL intensity and scattering contrast are directly correlated, as seen in Fig.~\ref{Fig_PL-scatt} for flakes of different thickness. The scattering contrast for treated MLG does not scale linearly with the number of layers, as in the case of pristine SLG~\cite{CasiraghiNL}. The MLG spectrum
only features negative contrast, while the positive contribution below 580nm observed for PLG is absent. Scattering spectra from treated MLG can be represented by a superposition of treated and pristine SLG.

Oxygen plasma etching of graphite proceeds layer-by-layer~\cite{alassadi93}. Thus, in our case only the topmost layer is affected. The absence of PL in MLG means that emission from the topmost layer is quenched by subjacent un-treated layers. This opens the possibility of engineering sandwiched hybrid structures consisting of PLG and a variable number of pristine graphene layers.

Oxygen plasma etching is expected to yield CO and CO$_2$, by successively removing carbon atoms. Etching of graphite occurs both in the basal plane and at defects\cite{eggito}. The latter is consistent with our observation of point-like PL features for short treatment times (Fig.~\ref{Fig_image}(a)).

It would be tempting to interpret the PL emission as coming from electron confinement in sp$^2$ islands with an average size of $\sim$1nm, as indicated by Raman spectroscopy. Indeed, since electrons in graphene behave as massless particles, energy quantization due to confinement is expected to open a gap $\delta E \approx v_F h /2d\approx2 eV nm/d$. The resulting quantum confined energy for a quantum disk of diameter d=2nm is 1eV. The
observed emission energy distribution translates into a diameter distribution ranging from 0.94 to 1.29nm, in agreement with the Raman estimation. In this case the large spectral width of the PL signal, $\sim$0.5eV, could result from a superposition of overlapping bands with narrow linewidth centered at different size-controlled (or quantum confined) energies, corresponding to heterogeneous broadening. Then, the optical properties of the PLG  would resemble those of pi-conjugated polymer films, where a distribution of conjugation lengths translates into a strong in-homogenously broadened density of states\cite{Bassler}. At room temperature, laser irradiation in the red would lead to selective excitation of a subset of quantum confined states. Then, spectral hole burning, i.~e. the selective photobleaching of this subset of homogenously broadened lines, should be possible. This bleaching could be a photochemical modification or even a complete removal of the absorbing subset. As a result we would observe a spectral hole, i.e. the subset absorbing a certain color would not contribute to PL. Fig.5 plots the ratio of PL measured before ($I_{PL}$) and after ($I_{PL,bleached}$) exposure to high power ($>$600$\mu$W) pulsed laser light at 647nm. The PL is measured at 530nm for low excitation power ($\sim10\mu$W). No spectral hole is observed in the detected spectral range, as would be expected for a heterogeneous ensemble of narrow bandwidth emitters. Instead, only an irreversible and uniform reduction of PL intensity occurs. For other bleaching energies in the red spectral range (760,800nm) the same uniform decrease is observed, while in the blue (473,514nm) the PL slightly shifts to shorter wavelength (see Fig.1c).
\begin{figure}
\centerline{\includegraphics [width=80mm]{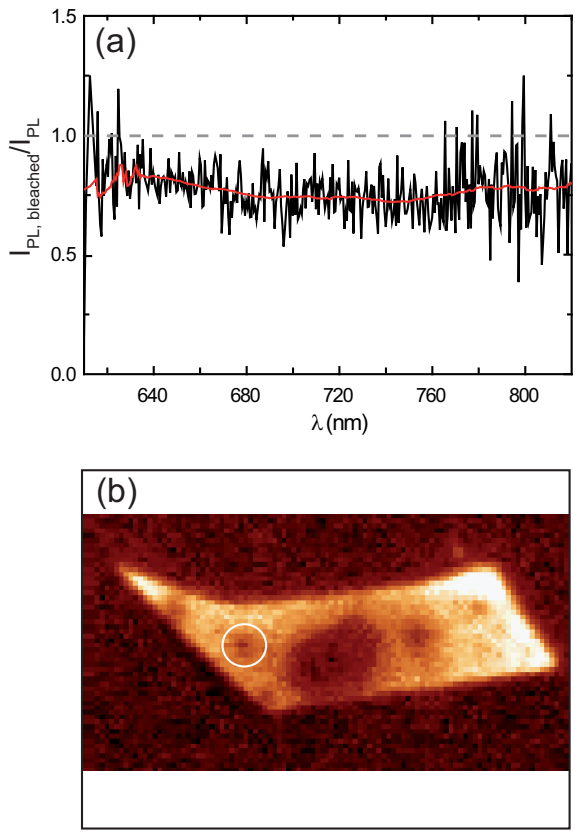}}
\caption{\label{FigS1}a)PL Intensity ratio for the area indicated by the circle in b).  }
\end{figure}

Thus, we find that the observed large spectral width of 0.5eV reflects mainly homogenous broadening, uniform across the PLG sheet. This is supported by the absence of spectral diffusion in the time-resolved data, expected for heterogeneous films\cite{Bassler} (see Methods). If PL would indeed result from quantum confined states~\cite{kikkawa}, size-related heterogenoeus broadening would need to be far smaller, probably below 0.1eV, requiring a very narrow size distribution of $\sim\pm $0.04nm, instead of the $\sim \pm $0.18nm needed for the 0.5eV broadening. Since oxidation is expected to occur at different lattice sites and configurations, such high degree of ordering would seem unreasonable. Moreover, while for increasing oxidation times a successive decrease of the effective size distribution would be expected, the spectral characteristics of the PL emission remain nearly constant. In conclusion, although the identification of $L_a$ as the quantum confinement length of massless electrons would be tempting, we rather assign the observed PL to CO-related localized electronic states at the oxidation sites.

In summary, we have shown that spatially uniform PL can be induced in single-layer graphene on substrates by selective plasma oxidiation. Remarkably, bi- and mutli-layer flakes remain non-luminescent, while their elastic scattering spectra indicate the formation of sandwich-like structures containing unetched layers. The resulting photoluminescent material could pave the way towards graphene-based optoelectronics.

ACF, KSN and AKG thank the Royal Society and the European Research Council (grants NANOPOTS and GRAPHENE). AH the Deutsche Forschungsgemeinschaft (DFG-grant HA4405/4-1)
and Nanosystems Initiative Munich (NIM)

\section*{Methods}
\subsection{Raman Background}
Raman spectroscopy is a fast and non-destructive method for the characterization of carbons. These show common features in the 800-2000 cm$^{-2}$ region: the G and D peaks. The G peak corresponds to the E$_{2g}$ phonon at the Brillouin zone centre. The D peak is due to the breathing modes of sp$^2$ rings and requires a defect for its activation\cite{Ferrari00,thomsen,tk}. It comes from TO phonons around \textbf{K}\cite{Ferrari00,tk}, is active by double resonance (DR)\cite{thomsen} and is strongly dispersive with excitation energy due to a Kohn Anomaly at \textbf{K}\cite{piscanec}. The 2D peak is the second order of the D peak. This is a single band in monolayer graphene, whereas it splits in four in bi-layer graphene, reflecting the evolution of the band structure\cite{ACFRaman}. The 2D peak is always seen, even when no D peak is present, since no defects are required for the activation of two phonons with the same momentum, one backscattering from the other. DR can also happen as intra-valley process, i.e. connecting two points belonging to the same cone around \textbf{K} or \textbf{K'}. This gives rise to the D' peak,$\sim$1620cm$^{-1}$ in defected graphite. The 2D' is the second order of the D' peak.

Ref.\cite{tk}, noted that I(D)/I(G) varies inversely with the cluster size $L_a$ in poly and nano crystalline graphites: I(D)/I(G)=C($\lambda$)/L$_a$, where C(514.5nm) $\sim$ 4.4nm from Refs.\cite{tk,knight}. The original idea was to link I(D) to phonon confinement. The intensity of the non-allowed D peak would be ruled by the defect-induced lifting of the Raman fundamental selection rule. Assuming that graphite becomes uniformly nano-crystalline, the D peak evolution can be estimated using Heisenberg indetermination principle: I(D)$\propto\Delta$q, with  $\Delta$q$\Delta$x$\propto$$\hbar$ and $\Delta$x$\sim$L$_a$. We now know that the D peak activation is due to DR and not to phonon confinement. However, also in this case, the higher the number of defects, the higher the chance of phonon-defect scattering and, thus, the higher I(D). Again, since the G peak is not defect-activated, even within DR, one can expect TK to hold. Now L$_a$ is an average inter-defect distance, instead of a cluster size. This is a very simple picture, which has proven effective to compare graphitic samples for increasing disorder. However, we note that a complete theory for the D and G Raman intensity and their dependence on the number of defects is still lacking.
\subsection{PL Transients}
\begin{figure}
\centerline{\includegraphics [width=80mm]{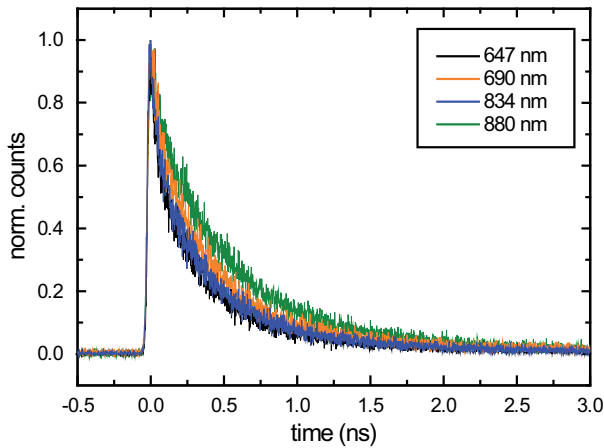}}
\caption{\label{FigS2}PL transients measured at four different detection energies after excitation at 530 nm. These are nearly uniform across the complete spectrum, indicating that spectral diffusion due to energy migration, typical for heterogeneously broadened systems, is absent.}
\end{figure}

Figure 6 plots the PL transients measured at four different detection energies after excitation at 530 nm. All can be modeled by a tri-exponential decay with decay constants of 30, 250 and 1000 ps, with different relative contributions. Remarkably, there is no direct correlation between emission energy and decay dynamics, as would be expected for heterogeneously broadened systems. In this case, spectral diffusion due to energy migration would lead to faster decay in the blue spectral range and a delayed signal rise on the same time scale in the red\cite{Bassler}. The decay traces can also be modeled using a stretched-exponential model function.

\end{document}